\documentclass[conference]{IEEEtran}
\IEEEoverridecommandlockouts
\usepackage{cite}
\usepackage{amsmath,amssymb,amsfonts}
\usepackage{algorithmic}
\usepackage{graphicx}
\usepackage{textcomp}
\usepackage{xcolor}
\usepackage{booktabs, multirow,array}
\usepackage{float}
\usepackage{subfig}
\usepackage{upgreek}

\def\BibTeX{{\rm B\kern-.05em{\sc i\kern-.025em b}\kern-.08em
    T\kern-.1667em\lower.7ex\hbox{E}\kern-.125emX}}
\begin{document}

\title{A Novel Waveform Design for OFDM-Based\\Joint Sensing and Communication System
%hould not be used}
}
\makeatletter
\newcommand{\linebreakand}{%
  \end{@IEEEauthorhalign}
  \hfill\mbox{}\par
  \mbox{}\hfill\begin{@IEEEauthorhalign}
}
\makeatother
\author{\IEEEauthorblockN{Yi Geng}
	\IEEEauthorblockA{\textit{Cictmobile, China} \\
		gengyi@cictmobile.com}
}
\maketitle

\begin{abstract}
The dominating waveform in 5G is orthogonal frequency division multiplexing (OFDM). OFDM will remain a promising waveform candidate for joint communication and sensing (JCAS) in 6G since OFDM can provide excellent data transmission capability and accurate sensing information. This paper proposes a novel OFDM-based diagonal waveform structure and corresponding signal processing algorithm. This approach allocates the sensing signals along the diagonal of the time-frequency resource block. Therefore, the sensing signals in a linear structure span both the frequency and time domains. The range and velocity of the object can be estimated simultaneously by applying 1D-discrete Fourier transform (DFT) to the diagonal sensing signals. Compared to the conventional \mbox{2D-DFT} OFDM radar algorithm, the computational complexity of the proposed algorithm is low. In addition, the sensing overhead can be substantially reduced. The performance of the proposed waveform is evaluated using simulation and analysis of results.
\end{abstract}


\section{Introduction}
Sixth generation (6G) will not be only about communication. Joint communication and sensing (JCAS), which will enable a myriad of new use cases, is a significant area of interest within 6G research. To achieve a favorable trade-off between communication and sensing, the waveforms for 6G need to be designed for simultaneous communication and sensing \cite{b1}. Orthogonal frequency division multiplexing (OFDM) is a promising waveform candidate for JCAS systems. Compared to other JCAS waveform candidates, e.g., frequency modulated continuous wave (FMCW), OFDM waveform naturally supports MIMO processing and has excellent communication performance \cite{b2}. In terms of sensing performance, OFDM is well-suited for range and velocity estimation. For example, an OFDM-based periodogram algorithm can obtain range-velocity estimation by applying 2D-discrete Fourier transform (DFT) to signals in modulation symbol domain \cite{b3}\cite{b4}. On the other hand, OFDM waveform avoids extra hardware complexity and costs compared to dual-waveform systems (e.g., time-multiplexing of OFDM and FMCW) \cite{b3}.

The paper is structured as follows. Section~II gives an example of how to design the sensing signal structure according to the requirements of a traffic monitoring scenario. Section~III provides an overview of the periodogram algorithm. In Section~IV, we propose a novel diagonal waveform structure and corresponding signal processing algorithm. Section~V concludes the paper.
\begin{table}[t]
	\caption{Exemplary KPIs for traffic monitoring}
	\begin{center}
		\begin{tabular}{ccc}
			\hline
			Requirement & Symbol & Value \\
			\hline
		Range resolution & $\Delta{R}$ & 0.4~m\\
        Velocity resolution & $\Delta{v}$ & 0.2~m/s\\
        Maximum detection range & $R_{\text{max}}$ & 150~m\\
        Maximum detection velocity & $v_{\text{max}}$ & 90~m/s\\
        \hline
		\end{tabular}
	\end{center}
	\label{tab_1}
\end{table}
\begin{table}[t]
	\caption{OFDM system parameters under the condition of satisfying the KPIs in Table~I\label{tab:table2}}
	\begin{center}
		\begin{tabular}{ccc}
			\hline
		System parameter & Symbol & Value\\
\hline
Carrier frequency & $f_\text{c}$ & 28 GHz\\
Bandwidth & $B$ & 400~MHz\\
Subcarrier spacing & SCS ($\Delta{f}$) & 120 KHz\\
Total subcarriers & $N_\text{c}$ & 3360\\
OFDM symbol duration & $T_{\text{sym}}$ & 8.92 $\upmu{s}$\\
OFDM slot duration & $T_\text{s}$ & 0.125 ms\\
Time-domain duration & $T_\text{b}$ & 240$T_\text{s}$ (30~ms)\\
Total symbols in $T_\text{b}$ & $N_\text{{sym}}$ & 3360\\
Comb size in frequency domain & $C_\text{f}$ & 7$\Delta{f}$\\
Comb size in time domain & $C_\text{t}$ & 7$T_{\text{sym}}$\\
Sensing signals in frequency domain & $N_\text{f}$ & 480\\
Sensing signals in time domain & $N_\text{t}$ & 480\\
Sensing signals along diagonal & $N$ & 480\\
\hline
		\end{tabular}
	\end{center}
	\label{tab_2}
\end{table}
\begin{figure}[t]
	\centering
	\subfloat[A comb structure of sensing signals]{\label{fig:1a}\includegraphics[width=0.8\columnwidth]{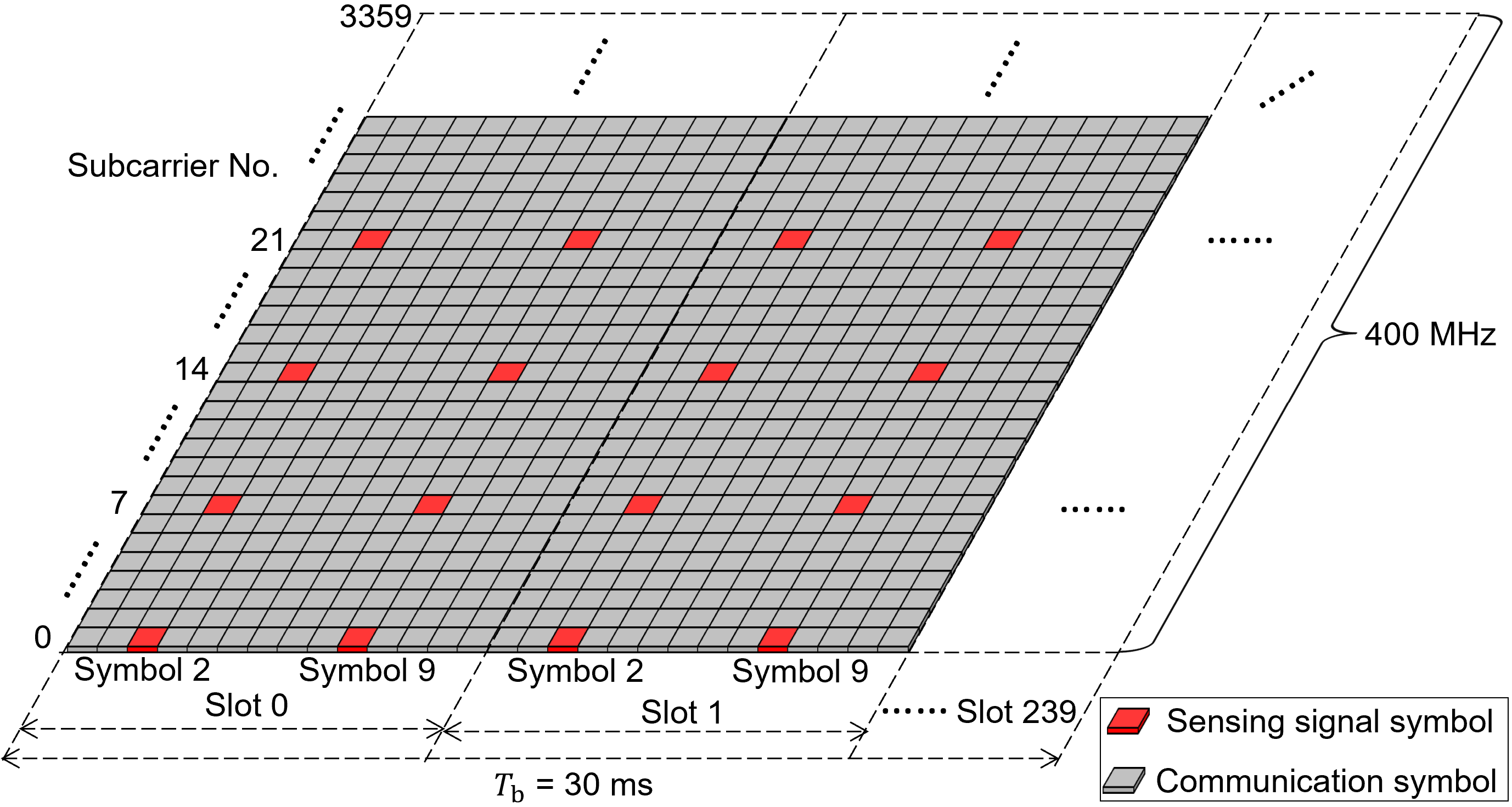}}\quad
	\subfloat[Consecutively transmitted blocks of sensing signals]{\label{fig:1b}\includegraphics[width=0.8\columnwidth]{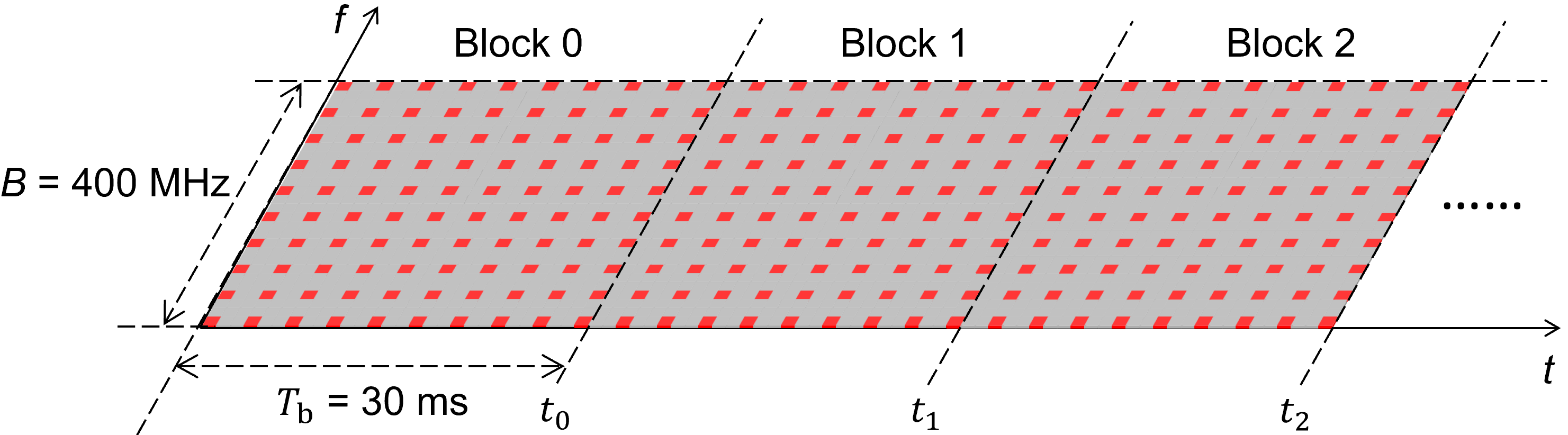}}\\
	\caption{A comb structure of sensing signals to meet the KPIs in Table~I.}
	\label{fig_1}
\end{figure}

\begin{figure}[t]
	\centerline{\includegraphics[width=0.8\linewidth, height=10cm, keepaspectratio]{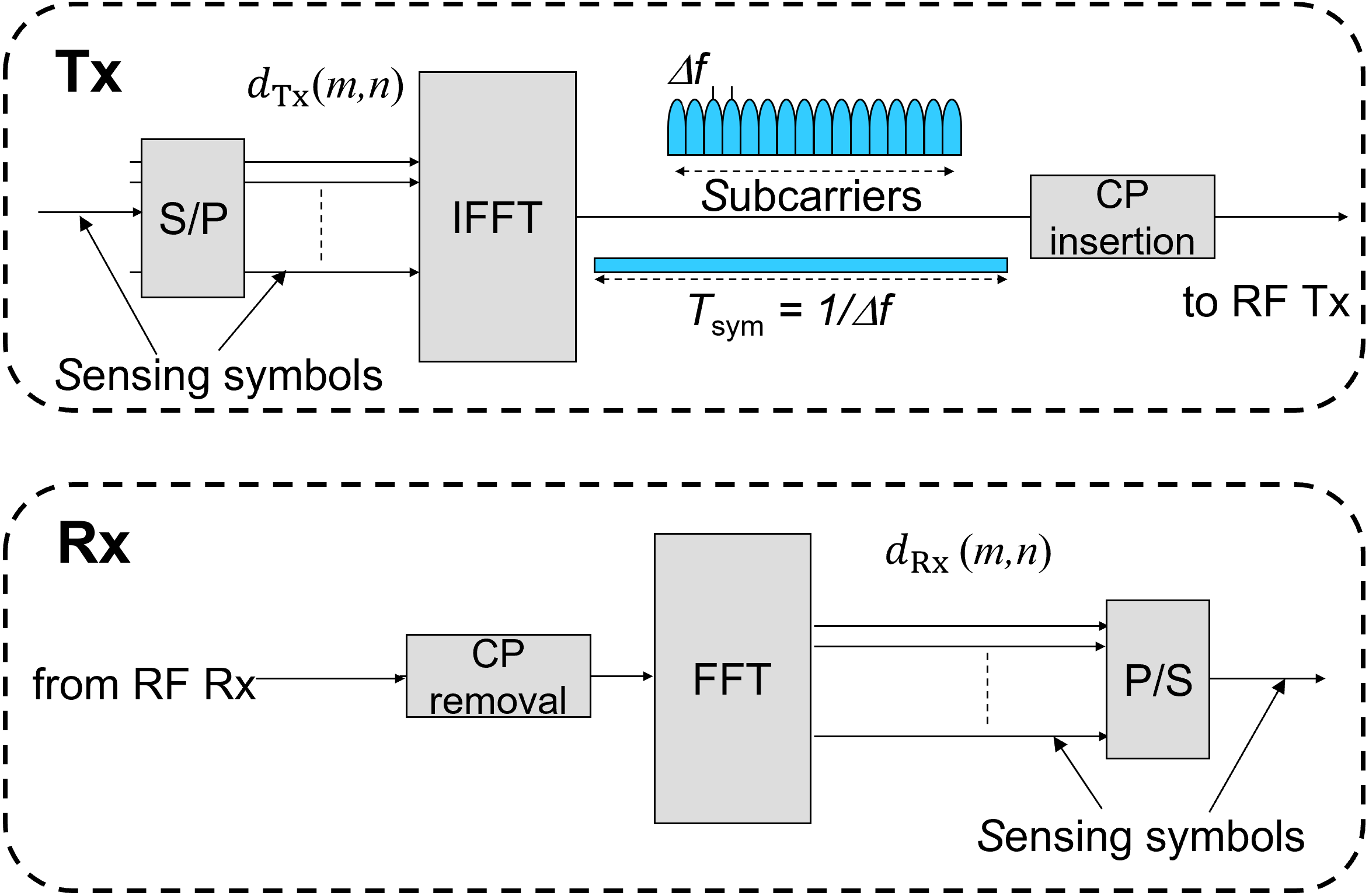}}
	\caption{Tx and Rx scheme of OFDM-based JCAS system.}
	\label{fig_2}
\end{figure}

\section{Waveform Design for JCAS System}
The waveform design of an OFDM-based JCAS system depends on the key performance indicators (KPIs) requested by the sensing applications, such as range resolution $\Delta{R}$, velocity resolution $\Delta{v}$, maximum detection range $R_{\text{max}}$, and maximum detection velocity $v_{\text{max}}$. For example, an OFDM-based JCAS system at 28~GHz carrier frequency ($f_\text{c}$) with 120~kHz subcarrier spacing (SCS), which is deployed for traffic monitoring and communication simultaneously, is designed to meet the KPIs tabulated in Table~I. To realize such a system, the OFDM system parameters that will be used in this paper are given in Table~II. A waveform structure to meet the KPIs in Table~I is shown in Fig.~1(a). To reduce the sensing overhead, the sensing subcarriers are assigned in a comb structure. $N_\text{f}$ sensing signals are uniformly distributed at an interval of comb size $C_\text{f}$ = 7$\Delta{f}$ within the band $B$. Note that the $N_\text{f}$ sensing signals occupy the full bandwidth $B$. Hence no deterioration of range resolution occurs \cite{b5}. The range resolution of this waveform $\Delta{R}$ is given by
\begin{equation}\label{eqn_1}
	\Delta{R}=\frac{\text{c}}{2B},
\end{equation}
where c is the speed of light.

The maximum unambiguous detection range of this interleaved scheme $R_{\text{max}}$ is reduced by a factor $\frac{N_\text{c}}{N_\text{f}}$ compared to the maximum unambiguous range with a classical contiguous subcarrier allocation. $R_{\text{max}}$ is given by
\begin{equation}\label{eqn_2}
	R_{\text{max}}=\frac{\text{c}N_\text{f}}{2\Delta{f}N_\text{c}}.
\end{equation}

Similarly, in the time domain, $N_\text{t}$ sensing signals are uniformly distributed at an interval of comb size $C_\text{t}$ = 7$T_{\text{sym}}$ within 240 OFDM slots (30~ms). The sensing signals are transmitted at symbol~2 and symbol~9 of each slot. The velocity resolution of this waveform $\Delta{v}$ is thus
\begin{equation}\label{eqn_3}
	\Delta{v}=\frac{\text{c}}{2f_{\text{c}}T_{\text{b}}}.
\end{equation}

The maximum unambiguous velocity $v_{\text{max}}$ can be given by
\begin{equation}\label{eqn_4}
	v_{\text{max}}=\frac{\text{c}\Delta{f}N_\text{t}}{2f_{\text{c}}N_\text{sym}}.
\end{equation}

According to \eqref{eqn_1}-\eqref{eqn_4}, a sensing system with the parameters in Table~II would be suitable to support the KPIs listed in Table~I. One sensing block illustrated in Fig.~1(a) can be used to derive one range-velocity estimate. The sensing blocks can be transmitted consecutively to track the objects with an update rate of 33.3~Hz, as illustrated in Fig.~1(b).
\section{OFDM Range-Doppler Processing in the Modulation Symbol Domain}
To date, various OFDM-based radar algorithms have been developed. In this section, a periodogram algorithm in the ``modulation symbol" domain \cite{b4} is presented. Fig.~2 illustrates the transmitting and receiving block diagram of an OFDM-based JCAS system using the algorithm in the cited work \cite{b4} and using the waveform shown in Fig.~1. At Tx, sensing signals are converted from serial to parallel. Each parallel symbol stream modulates a 120~kHz subcarrier. Over 30~ms, a 480$\times$480 modulation symbol matrix $\mathbf{D}_{\text{Tx}}(m,n)$ is processed by Inverse Fast Fourier Transform (IFFT). Each row of $\mathbf{D}_{\text{Tx}}(m,n)$ represents a vector carrying Doppler information obtained from a sensing subcarrier. The indices of sensing subcarriers $m$ in the frequency domain range from 0 to $N_\text{f}-1$. Each column of $\mathbf{D}_{\text{Tx}}(m,n)$ represents a vector carrying range information obtained from a sensing symbol. The indices of sensing symbols $n$ range from 0 to $N_\text{t}-1$. After IFFT processing, CP insertion, and digital-to-analog conversion, the matrix $\mathbf{D}_{\text{Tx}}(m,n)$ is transmitted in the air. The objects in the detection range affect the propagation of $\mathbf{D}_{\text{Tx}}(m,n)$. Therefore, the received modulation symbol matrix $\mathbf{D}_{\text{Rx}}(m,n)$ is the combination of $\mathbf{D}_{\text{Tx}}(m,n)$ and information of the objects. The user data carried by the sensing signals are eliminated by element-wise division between $\mathbf{D}_{\text{Rx}}(m,n)$ and $\mathbf{D}_{\text{Tx}}(m,n)$, yielding
\begin{equation}\label{eqn_5}
	\mathbf{D}(m,n)=\frac{\mathbf{D}_{\text{Rx}}(m,n)}{\mathbf{D}_{\text{Tx}}(m,n)} = y_\text{R}(m)\otimes{y_\text{D}(n)},
\end{equation}
where
\begin{equation}\label{eqn_6}
	y_\text{R}(m) = \text{exp}({\frac{-\text{j}4\pi\Delta{f}Rm}{\text{c}}}), m=0, \cdots,N_\text{f}-1
\end{equation}
\begin{equation}\label{eqn_7}
	y_\text{D}(n) = \text{exp}({\frac{\text{j}4\pi{T_{\text{sym}}}{f_{\text{c}}}vn}{\text{c}}}), n=0, \cdots,N_\text{t}-1
\end{equation}
where $\mathbf{D}(m,n)$ is a $N_\text{f}\times{N_\text{t}}$ matrix, the operator $\otimes$ denotes dyadic product, $y_\text{R}(m)$ and $y_\text{D}(n)$ are $N_\text{f}\times{1}$ vector and $1\times{N_\text{t}}$ vector, respectively. $R$ is the range between the JCAS antenna and the object, $v$ is the radial velocity of the object.

The linear phase shifts of $y_\text{R}(m)$ and $y_\text{D}(n)$ carry the range and Doppler information of the object. When a $\mathbf{D}(m,n)$ is extracted from a sensing block, inverse discrete Fourier transform (IDFT) and DFT are performed in the frequency domain and time domain, respectively, to derive the range $R$ as well as the velocity $v$ of the object \cite{b6}\cite{b7},
\begin{multline}\label{eqn_8}
	Y_r(p) = \text{IDFT}(y_\text{R}(m))=\frac{1}{N_\text{f}}\sum_{m=0}^{N_\text{f}-1}y_\text{R}(m)\text{exp}({\frac{\text{j}2\pi{m}p}{N_\text{f}}})\\
	=\frac{1}{N_\text{f}}\sum_{m=0}^{N_\text{f}-1}\text{exp}({\frac{-\text{j}4\pi\Delta{f}Rm}{\text{c}}})\text{exp}({\frac{\text{j}2\pi{m}p}{N_\text{f}}}),\\
	p=0, \cdots,N_\text{f}-1
\end{multline}
\begin{multline}\label{eqn_9}
	Y_v(q) = \text{DFT}(y_\text{D}(n))=\sum_{n=0}^{N_\text{t}-1}y_\text{D}(n)\text{exp}({-\frac{\text{j}2\pi{n}q}{N_\text{t}}})\\
	=\sum_{n=0}^{N_\text{t}-1}\text{exp}({\frac{\text{j}4\pi{T_{\text{sym}}}{f_{\text{c}}}vn}{\text{c}}})\text{exp}({-\frac{\text{j}2\pi{n}q}{N_\text{t}}}),\\
	q=0, \cdots,N_\text{t}-1
\end{multline}

By performing IDFT and DFT, the phase shifts of $y_\text{R}(m)$ and $y_\text{D}(n)$ are transformed from the time-frequency domain to spectral peaks in the delay domain and Doppler domain \cite{b8}. Range $R$ can be calculated by the IDFT bin index $p$ in the delay domain where a peak occurs. Similarly, velocity $v$ can be obtained by the DFT bin index $q$ in the Doppler domain where a peak occurs. The range and velocity can be calculated by
\begin{equation}\label{eqn_10}
	R = \frac{\text{c}p_{\text{peak}}}{2\Delta{f}N_\text{f}},
\end{equation}
\begin{equation}\label{eqn_11}
	v = \frac{\text{c}q_{\text{peak}}}{2f_cT_{\text{sym}}N_\text{t}},
\end{equation}
where $p_{\text{peak}}$ is the bin index at peak location in the delay domain, $q_{\text{peak}}$ is the bin index at peak location in the Doppler domain.

Matrix $\mathbf{D}(m,n)$ in \eqref{eqn_5} can be rewritten as
\begin{equation}\label{eqn_12}
	\mathbf{D}(m,n)=
	\begin{pmatrix} 
		\text{D}(0,0) & \ldots & \text{D}(0,N_\text{t}-1)\\
		\vdots & \ddots & \vdots \\
		\text{D}(N_\text{f}-1,0) & \ldots & \text{D}(N_\text{f}-1,N_\text{t}-1)
	\end{pmatrix}.
\end{equation}

2D-DFT of $\mathbf{D}(m,n)$ can be performed to compute the range and velocity of the object. It can be implemented in two stages: column-by-column 1D-IDFTs of length $N_\text{f}$ are proceeded after row-by-row 1D-DFTs of length $N_\text{t}$ as shown in Fig.~3. Then, the 2D range-velocity periodogram can be obtained, giving an intuitive indication of the reflecting objects.
\begin{figure}[t]
	\centerline{\includegraphics[width=0.9\linewidth, height=10cm, keepaspectratio]{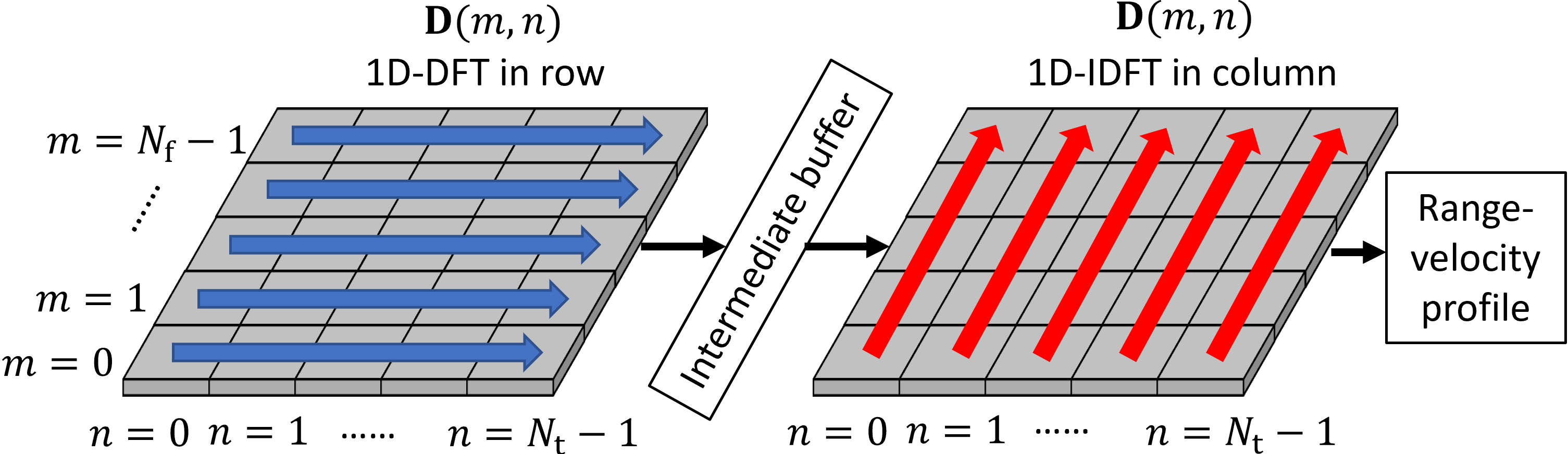}}
	\caption{Sensing signal processing of a matrix $\mathbf{D}(m,n)$ with application of 2D-DFT.}
	\label{fig_3}
\end{figure}
\section{A Novel Waveform Design and Sensing Signal Processing Algorithm}
\subsection{Challenges}
\begin{figure}[t]
	\centering
	\subfloat[A diagonal of sensing signals]{\label{fig:5a}\includegraphics[width=0.8\columnwidth]{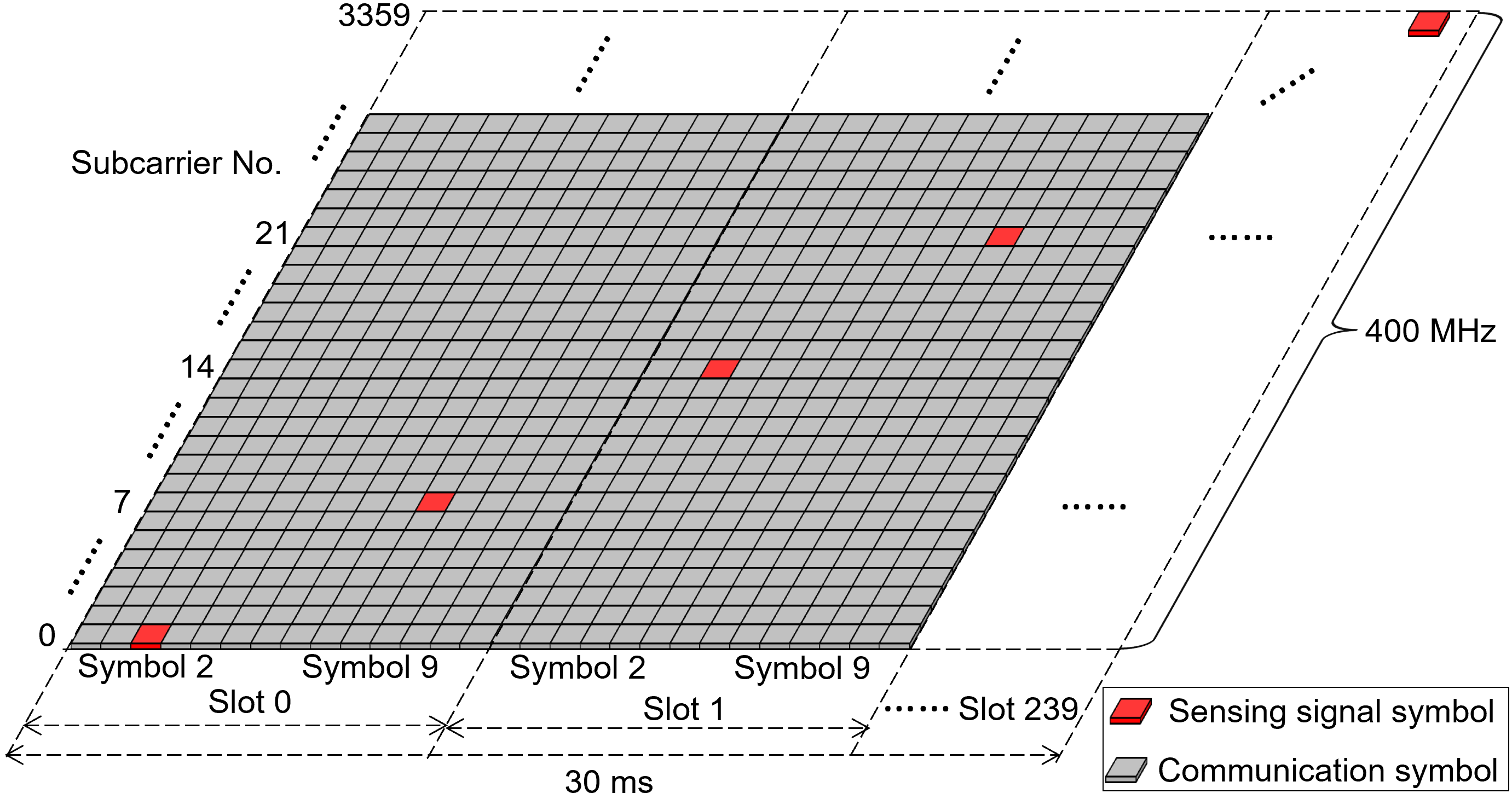}}\quad
	\subfloat[Consecutively transmitted diagonals of sensing signals]{\label{fig:5b}\includegraphics[width=0.8\columnwidth]{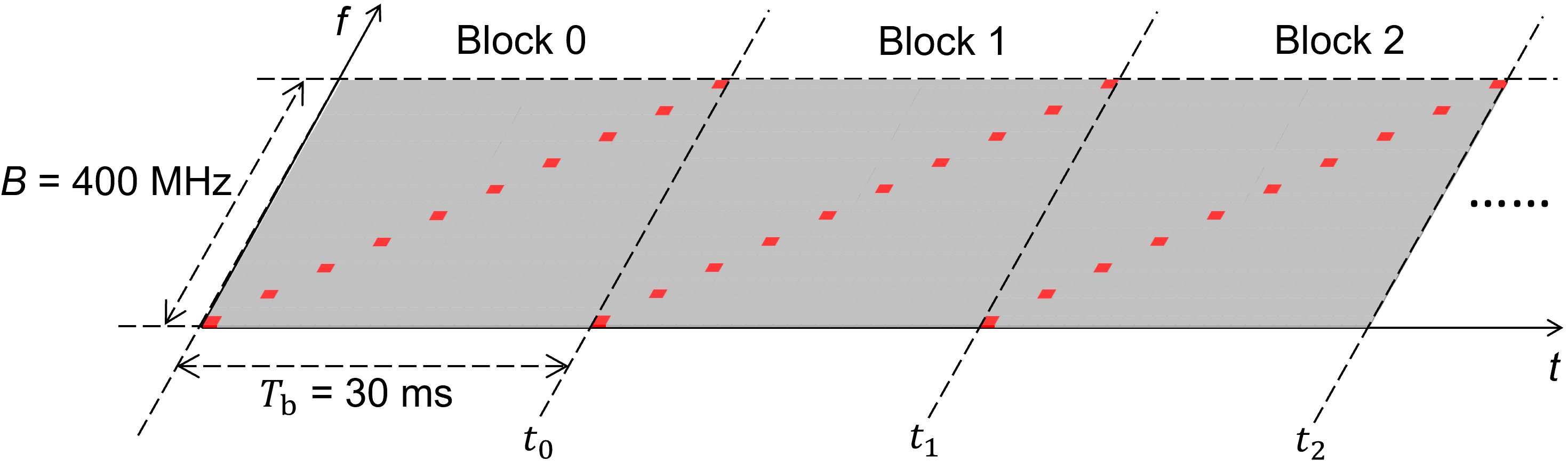}}\\
	\caption{A diagonal structure of sensing signals to meet the KPIs in Table~I.}
	\label{fig_4}
\end{figure}

Despite its performance and practicality, the 2D-DFT-based periodogram method suffers from several drawbacks. First, the 2D-DFT calculation is computationally expensive. The complexity of an $N$-point DFT is $\mathcal{O}(N^2)$. For example, to apply 2D-DFT to one matrix $\mathbf{D}(m,n)$ produced by the sensing block illustrated in Fig.~1(a), 960$\mathcal{O}(N^2)$ complex multiplications (480 DFTs in row and 480 IDFTs in column) are needed to generate one range-velocity estimate. Since the sensing system is ``unintelligent" and cannot predict the positions of objects in the detection range, it must scan the environment in each direction to localize and track the objects using narrow beams, leading to an extremely high computational complexity. Second, 2D-DFT is calculated in stages where all the results of the row-by-row \mbox{1D-DFTs} must be available before the column-by-column \mbox{1D-IDFTs} can be performed. Therefore, an additional intermediate cache is required, thus increasing hardware cost, especially for large DFT size. Finally, the sensing signals must be transmitted both in the time and frequency domain in a comb structure, which induces high sensing overhead.
\subsection{Method}
To tackle these challenges, a novel OFDM-based sensing waveform structure and corresponding signal processing algorithm are proposed. As shown in Fig.~4(a), a rectangular time-frequency block with a bandwidth of 400~MHz contains 3360 subcarriers in the frequency domain, and the block spans 30~ms in the time domain. Uniformly distributed sensing signals are allocated along the diagonal of the block. The time domain and the frequency domain contain the same number, $N$, of sensing signals. A diagonal of sensing signals produces a transmitted modulation symbol sequence $\mathbf{d}_{\text{Tx}}(k)$ of length $N$, spanning 240~slots in the time domain and 400~MHz in the frequency domain. The normalized modulation symbol vector $\mathbf{d}(k)$ can be obtained by the element-wise division between the received modulation symbol sequence $\mathbf{d}_{\text{Rx}}(k)$ and $\mathbf{d}_{\text{Tx}}(k)$, yielding
\begin{equation}\label{eqn_13}
	\mathbf{d}(k)=\frac{\mathbf{d}_{\text{Rx}}(k)}{\mathbf{d}_{\text{Tx}}(k)}=y_\text{R}(k)\otimes{y_\text{D}(k)},k=0, \cdots,N-1
\end{equation}
where
\begin{equation}\label{eqn_14}
	y_\text{R}(k) = \text{exp}({\frac{-\text{j}4\pi\Delta{f}Rk}{\text{c}}}),k=0, \cdots,N-1
\end{equation}
\begin{equation}\label{eqn_15}
	y_\text{D}(k) = \text{exp}({\frac{\text{j}4\pi{T_{\text{sym}}}{f_{\text{c}}}vk}{\text{c}}}),k=0, \cdots,N-1
\end{equation}

The range $R$ of the object causes the phase shifts of the individual elements of vector $y_\text{R}(k)$. The velocity $v$ of the object causes the phase shifts of the individual elements of vector $y_\text{D}(k)$. Since vector $y_\text{R}(k)$ and vector $y_\text{D}(k)$ contain the same number of elements and they are equally spaced in both the frequency domain and time domain, the range and velocity of the object can be obtained simultaneously by performing DFT of $\mathbf{d}(k)$, which yields
\begin{multline}\label{eqn_16}
	Y_{rv}(l)=\text{DFT}(\mathbf{d}(k))=
	\sum_{k=0}^{N-1}y_{\text{R}}(k)y_{\text{D}}(k)\text{exp}({-\frac{\text{j}2\pi{k}l}{N}}),\\
	l=0, \cdots,N-1
\end{multline}

\begin{figure}[t]
	\centerline{\includegraphics[width=0.8\linewidth, height=10cm, keepaspectratio]{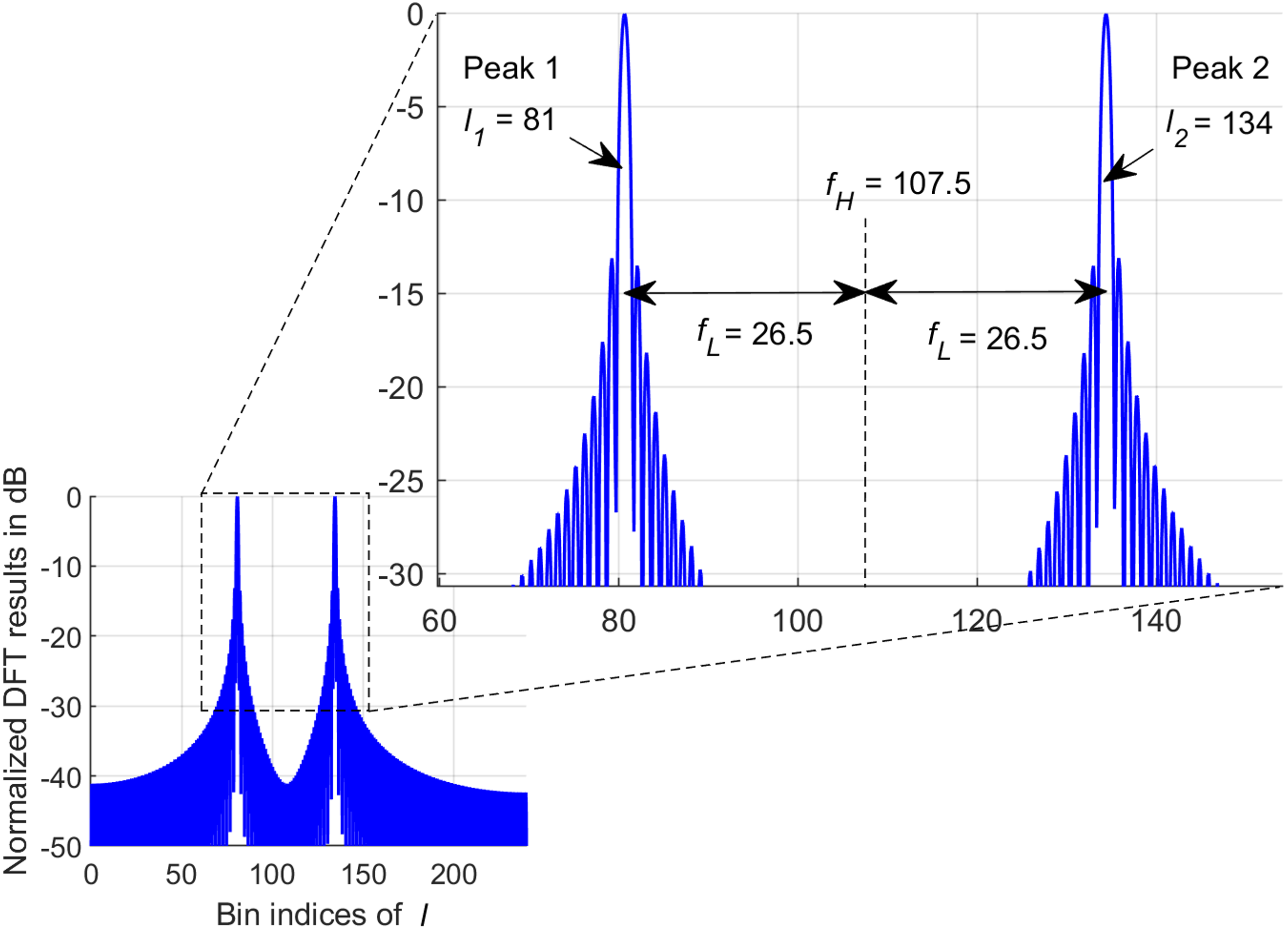}}
	\caption{The radar image of an object with range of 40~m and velocity of 5~m/s using the proposed algorithm.}
	\label{fig_5}
\end{figure}
Fig.~5 shows the normalized DFT result of a modulation symbol vector $\mathbf{d}(k)$, which is derived from the sensing signals reflected by a moving object with range \mbox{$R$ = 40~m} and velocity \mbox{$v$ = 5~m/s} at time $t_0$. The $x$-axis represents DFT bin indices $l$. The number of bins equals the DFT size $N$. The integer values on the $x$-axis correspond to the spectral frequencies sampled by the DFT. The DFT result shows a dual-peak-like profile. The bin indices at two peak locations are $l_1$ = 81 and $l_2$ = 134. Two spectral frequencies $f_\text{H}$ and $f_\text{L}$, which carry the range and Doppler information of the object, can be calculated by
\begin{equation}\label{eqn_17}
	f_\text{H} = \frac{l_1+l_2}{2},
\end{equation}
\begin{equation}\label{eqn_18}
	f_\text{L} = \frac{l_2-l_1}{2}.
\end{equation}

The DFT of $\mathbf{d}(k)$ translates the modulation signals in the time-frequency domain to two spectral peaks centered at $f_\text{H}$ and spaced at an interval of 2$f_\text{L}$ in the radar image. One of $f_\text{H}$ and $f_\text{L}$ contains range information, whereas the other contains velocity information. The range and velocity of the object can be extracted by
\begin{equation}\label{eqn_19}
	R = \frac{\text{c}f_\text{H}}{2\Delta{f}N_{\text{c}}}, v = \frac{
		\text{c}f_\text{L}}{2T_{\text{sym}}f_{\text{c}}N_{\text{c}}},
\end{equation}
or
\begin{equation}\label{eqn_20}
	R = \frac{\text{c}f_\text{L}}{2\Delta{f}N_{\text{c}}}, v = \frac{
		\text{c}f_\text{H}}{2T_{\text{sym}}f_{\text{c}}N_{\text{c}}}.
\end{equation}

As a side effect of the proposed algorithm, using \eqref{eqn_19} and \eqref{eqn_20} yields two rang-velocity estimates from one block of sensing signals. These two estimates include a correct range-velocity estimate and an incorrect range-velocity estimate. For example, two range-velocity estimates can be extracted from bin indices \mbox{$l_1$ = 81} and \mbox{$l_2$ = 134} shown in Fig.~5, referred to as estimate~A and estimate~B below.
\begin{itemize}
	\item{Estimate A: $R$ = 40~m and $v$ = 5~m/s}
	\item{Estimate B: $R$ = 10~m and $v$ = 20~m/s}
\end{itemize}

By using the proposed method, an object with \mbox{$R$ = 40~m} and \mbox{$v$ = 5~m/s} produces the same radar image as that of another object with \mbox{$R$ = 10~m} and \mbox{$v$ = 20~m/s}. The uncertainty is caused by the unlabeled bin indices $l_1$ and $l_2$ at peak locations. For spectral frequencies $f_\text{H}$ and $f_\text{L}$, it is uncertain which one is indicative of the range and which one indicates the velocity. These two estimates cannot be discerned using information from one sensing block. A multi-temporal data fusion method can be performed to filter out the incorrect estimate. The main idea of the multi-temporal data fusion method is as follows. Based on the two range-velocity estimates acquired in the recent past, the range-velocity estimate in the present can be predicted. The incorrect estimate can be identified by comparing the actual estimates obtained in the present with the predicted estimates.

\begin{figure*}[t]
	\centering
	\subfloat[Real radar image for a vehicle with $R$ = 40 m and constant $v$ = 5 m/s at time $t_0$, and predicted radar images at time $t_0$-$t_4$]{\label{fig:6a}\includegraphics[width=0.8\columnwidth]{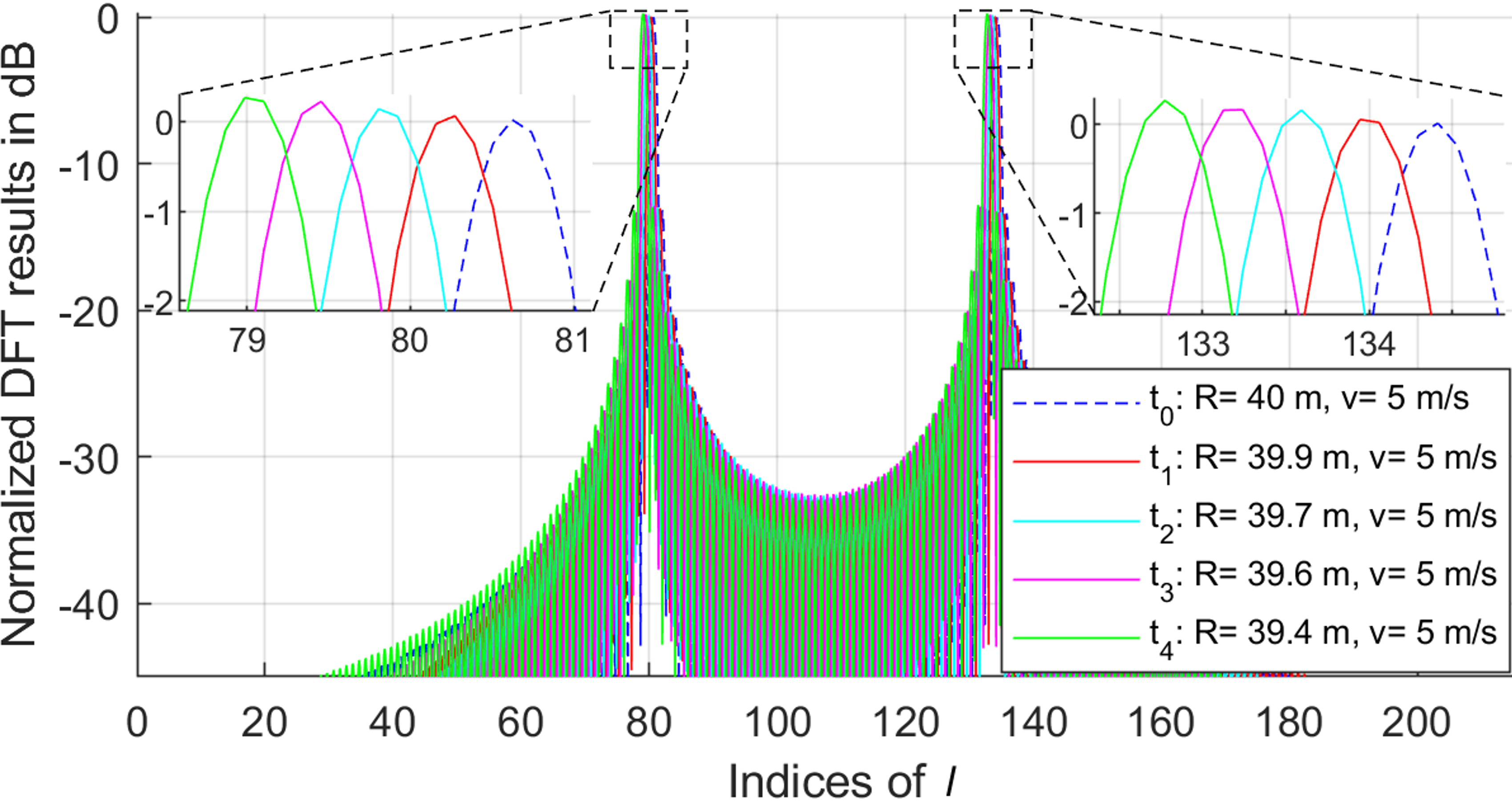}}\quad
	\subfloat[Real radar image for a vehicle with $R$ = 10 m and constant $v$ = 20 m/s at time $t_0$, and predicted radar images at time $t_0$-$t_4$]{\label{fig:6b}\includegraphics[width=0.8\columnwidth]{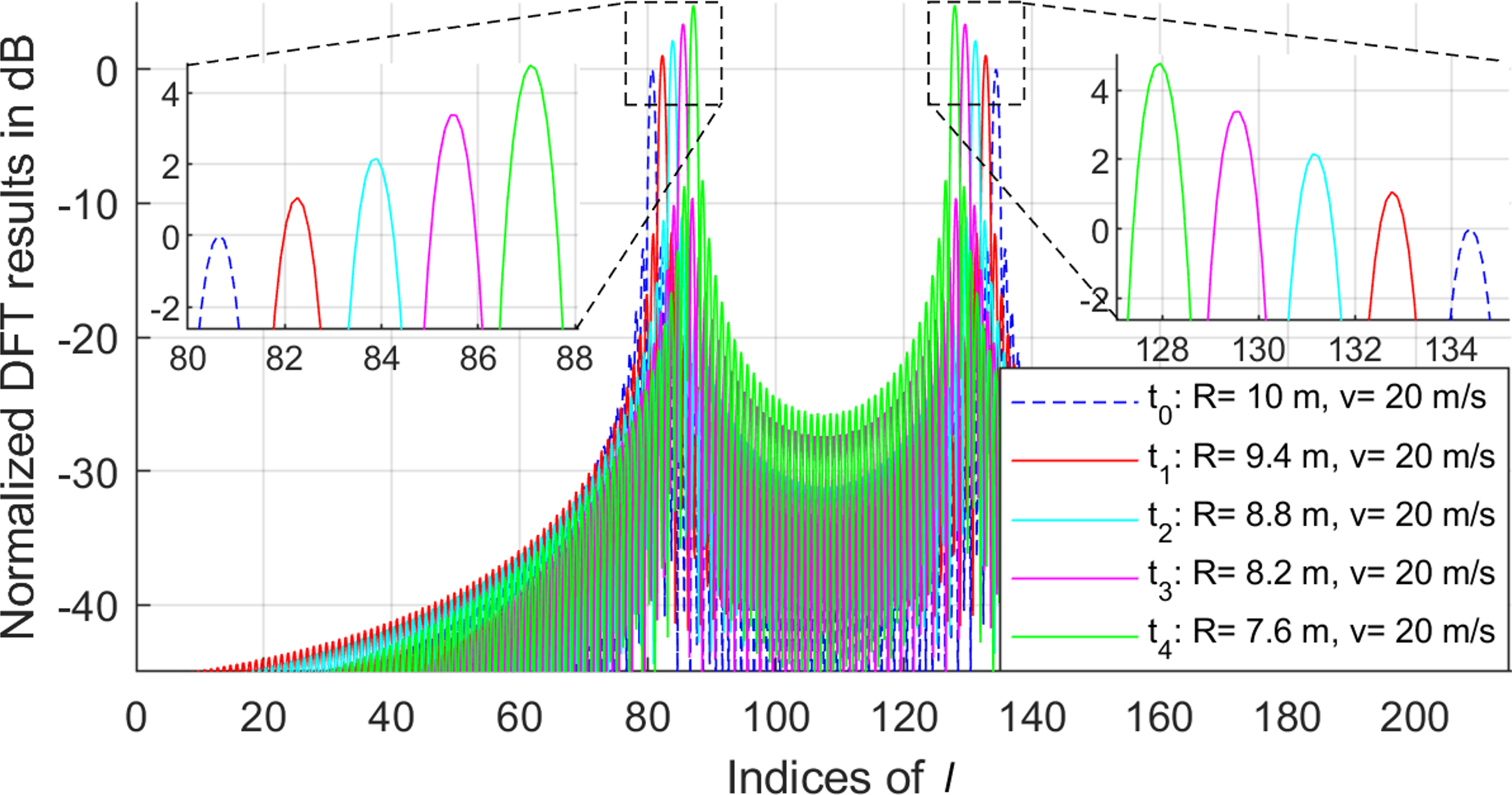}}\quad
	\subfloat[Real radar image for a vehicle with $R$ = 40 m, $v$ = 5 m/s and constant acceleration 5.4 m/s$^2$ at time $t_0$, and predicted radar images at time $t_0$-$t_4$]{\label{fig:6c}\includegraphics[width=0.8\columnwidth]{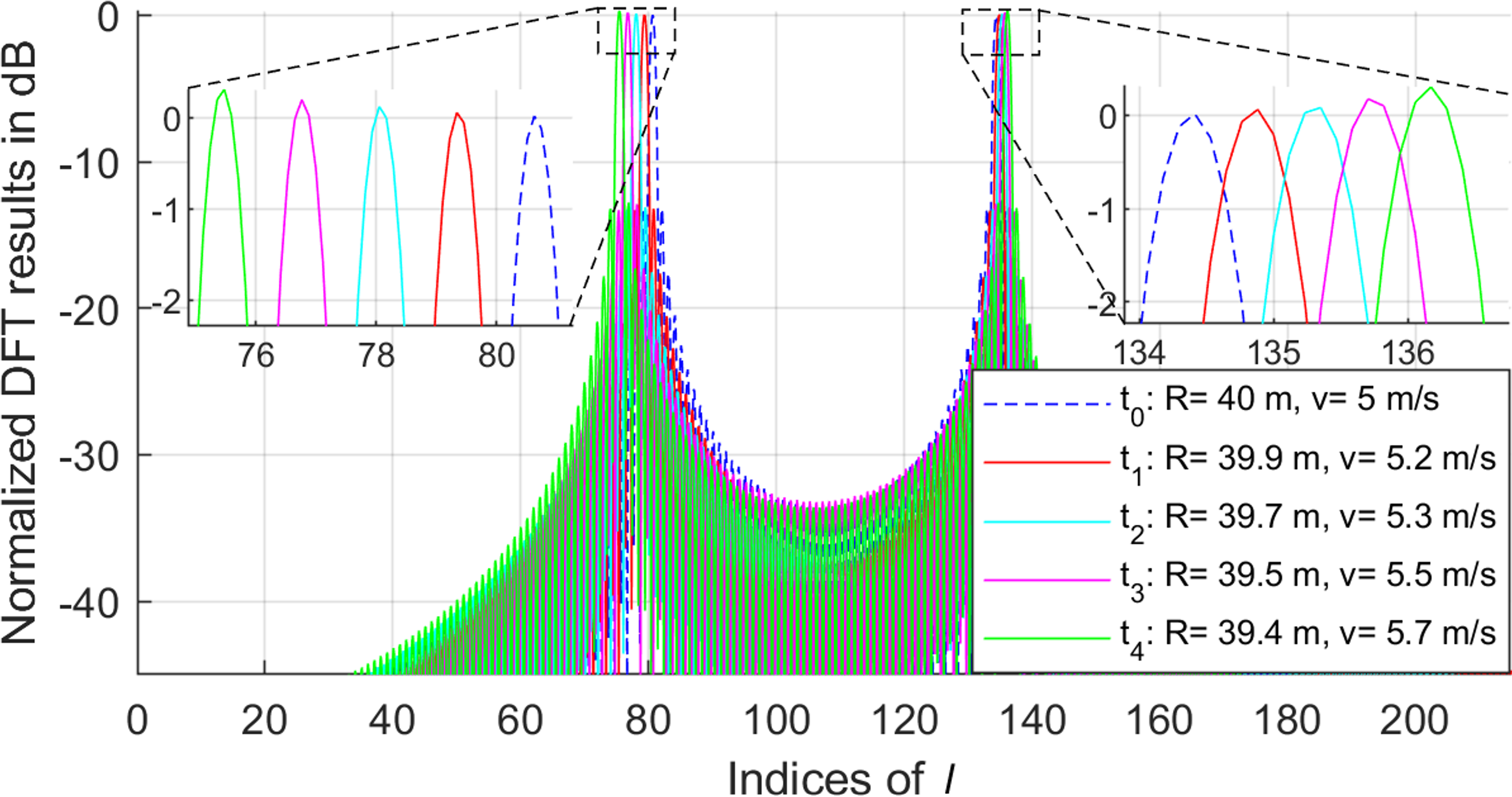}}\quad
	\subfloat[Real radar image for a vehicle with $R$ = 10 m, $v$ = 20 m/s and constant acceleration 5.4 m/s$^2$ at time $t_0$, and predicted radar images at time $t_0$-$t_4$]{\label{fig:6d}\includegraphics[width=0.8\columnwidth]{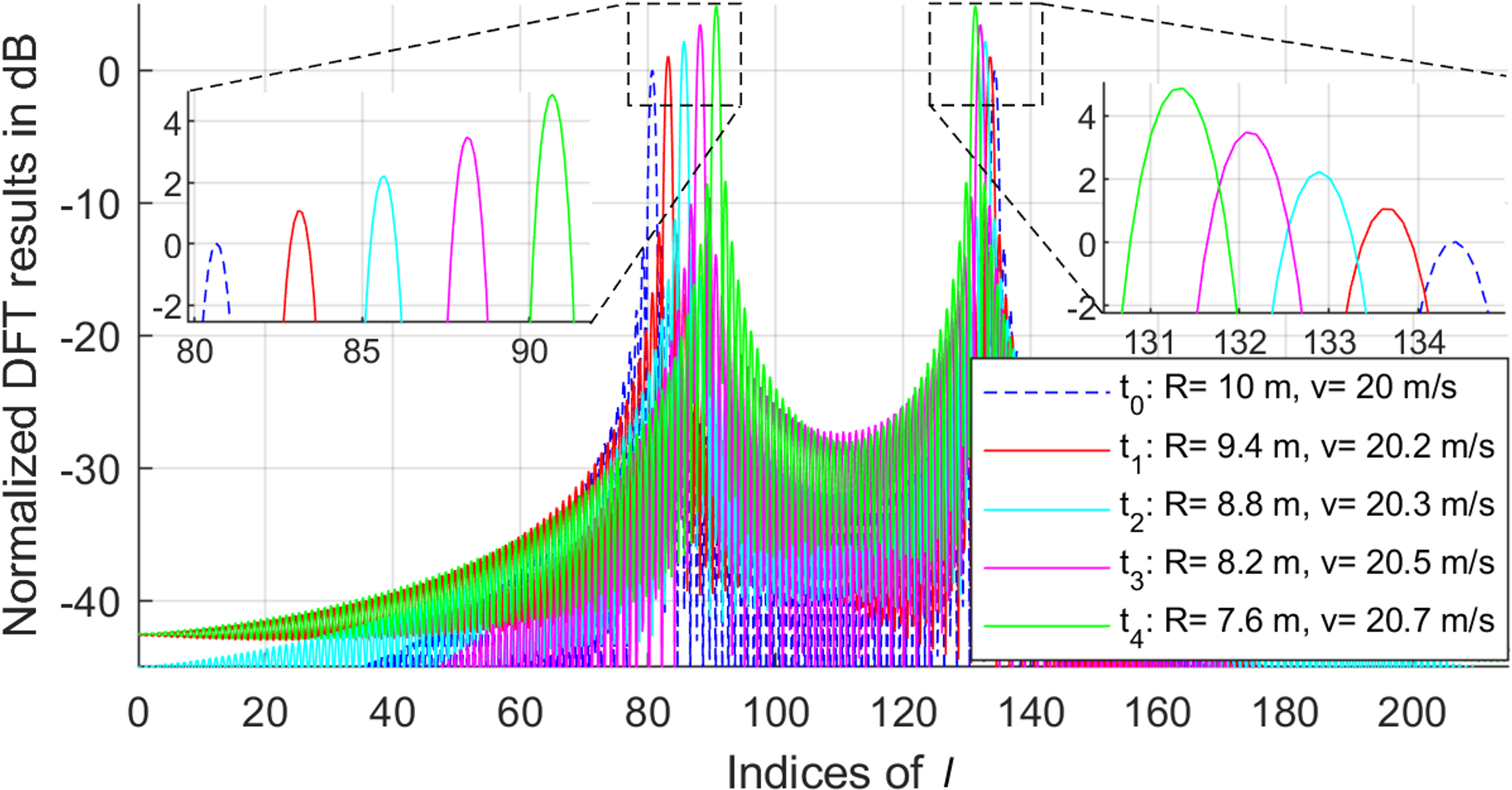}}\\
	\caption{Predicted radar images at time $t_1$-$t_4$ based on estimate~A and estimate~B obtained at time $t_0$.}
	\label{fig_6}
\end{figure*}

A traffic monitoring scenario is considered to show how the incorrect estimate is identified. We assume that the vehicles in the scenario comply with the linear motion model, which is popularly used in traffic research \cite{b9}. The linear motion model is a motion model where an object's velocity or acceleration is held constant within a span of time, provided that the time is sufficiently short. We assume the JCAS system has no prior range-velocity information on the vehicles. The vehicle's maximum acceleration ($a$) is assumed to be 5.4~m/s$^2$, corresponding to the acceleration from 0 to 60~mph in 5~seconds (acceleration of high-performance cars). The consecutive sensing blocks are transmitted at time 0~ms ($t_0$), 30~ms ($t_1$), 60~ms ($t_2$), 90~ms ($t_3$), 120~ms ($t_4$), and so on. For the linear motion model with constant velocity, the range and velocity of the vehicle at time $t_1$-$t_4$ can be predicted based on estimate~A and estimate~B obtained at time $t_0$. Similarly, the range and velocity of the vehicle at time $t_1$-$t_4$ can be obtained for the linear motion model with a constant acceleration of 5.4~m/s$^2$.

The amplitude of the spectral peak in the radar image is determined by range $R$ as it affects the received signal power $P_\text{R}$ reflected by the object. $P_\text{R}$ is formulated as
\begin{equation}\label{eqn_21}
	P_\text{R} = \frac{P_{\text{Tx}}G_{\text{Tx}}G_{\text{Rx}}\sigma\lambda^2}{(4\pi)^3R^4f_\text{c}^2},
\end{equation}
where $P_{\text{Tx}}$ is the transmitted power, $G_{\text{Tx}}$ is the Tx antenna gain, $G_{\text{Rx}}$ is the Rx antenna gain, $\sigma$ is the RCS of the object, $\lambda$ is the wavelength of the carrier.

\begin{table}[t]
	\caption{The predicted range and velocity of the vehicle at time $t_1$-$t_4$ based on estimate~A and estimate~B obtained at \mbox{time $t_0$}}
	\begin{center}
		\begin{tabular}{ccccc}
			\hline
			Estimate & $t$ (ms) & $R(t)$ (m) & $v(t)$ (m/s) & $A_\text{p}(t)$ (dB) \\
			\hline
			& $t_0$ = 0 & 40 & 5 & 0\\
			Estimate A & $t_1$ = 30 & 39.9 & 5 & 0.06\\
			with & $t_2$ = 60 & 39.7 & 5 & 0.13\\
			constant $v$ & $t_3$ = 90 & 39.6 & 5 & 0.19\\
			  & $t_4$ = 120 & 39.4 & 5 & 0.26\\
			\hline
			& $t_0$ = 0 & 10 & 20 & 0\\
Estimate B & $t_1$ = 30 & 9.4 & 20 & 1.07\\
with & $t_2$ = 60 & 8.8 & 20 & 2.22\\
constant $v$ & $t_3$ = 90 & 8.2 & 20 & 3.45\\
& $t_4$ = 120 & 7.6 & 20 & 4.77\\
            \hline
			& $t_0$ = 0 & 40 & 5 & 0\\
Estimate A & $t_1$ = 30 & 39.9 & 5.2 & 0.06\\
with & $t_2$ = 60 & 39.7 & 5.3 & 0.13\\
constant $a$ & $t_3$ = 90 & 39.5 & 5.5 & 0.2\\
& $t_4$ = 120 & 38.4 & 5.7 & 0.28\\
\hline
			& $t_0$ = 0 & 10 & 20 & 0\\
Estimate B & $t_1$ = 30 & 9.4 & 20.2 & 1.08\\
with & $t_2$ = 60 & 8.8 & 20.3 & 2.24\\
constant $a$ & $t_3$ = 90 & 8.2 & 20.5 & 3.49\\
& $t_4$ = 120 & 7.6 & 20.7 & 4.86\\
\hline
		\end{tabular}
	\end{center}
	\label{tab_3}
\end{table}

Based on estimate~A and estimate~B obtained at time~$t_0$ and the linear motion model, the predicted ranges $R(t)$, velocities $v(t)$ and normalized peak amplitudes $A_\text{p}(t)$ at time $t_1$-$t_4$ are tabulated in Table~III. $A_\text{p}(t_0)$ is normalized to 0~dB. The real radar images calculated at time $t_0$ and the predicted radar images at time $t_1$-$t_4$ in the future are depicted in Fig.~6. The blue dashed lines indicate the real radar images yielded by the sensing block transmitted at time $t_0$, while the predicted radar images at time $t_1$-$t_4$ are visualized with assorted colors. It can be observed that although a vehicle with \mbox{$R$ = 40~m} and \mbox{$v$ = 5~m/s} produces the same radar image as that of a vehicle with $R$ = 10~m and $v$ = 20~m/s at time $t_0$, the following radar images at time $t_1$-$t_4$ show different patterns. Fig.~6(a) shows minor peak shifts due to the elapsed time. Because constant velocity causes unchanged $f_{\text{L}}$, and the quite low velocity (5~m/s) causes minor range changes at time $t_1$-$t_4$. Hence, the changes in $f_{\text{H}}$ are less significant. The bin indices at peak locations at time $t_0$-$t_4$ are within 79-81 and 133-134, as shown in the zoomed-in sections in Fig.~6(a). Also, the peak amplitude increases slightly from 0~dB at $t_0$ to 0.26~dB at $t_4$. The minor range changes causes small received signal power changes, which, in turn, leads to slight peak amplitude fluctuation.

It can be observed that the radar images in Fig.~6(b) are pretty different from the radar images in Fig.~6(a). The constant velocity produces an unchanged spectral frequency $f_{\text{H}}$ in Fig.~6(b). Meanwhile, the high velocity (\mbox{$v$ = 20~m/s}) leads to significant changes in the range (major changes in spectral frequency $f_{\text{L}}$). The bin indices at peak locations at time $t_0$-$t_4$ are within 80-88 and 128-134, which are distinctly different from the bin indices in Fig.~6(a). Furthermore, due to the significant distance changes, the peak amplitude increases rapidly from 0~dB at $t_0$ to 4.77~dB at $t_4$. Therefore, for constant velocity, by comparing the bin indices and peak amplitudes from predicted radar images at time $t_1$-$t_4$ with the bin indices and peak amplitudes from actual radar images obtained at time $t_1$-$t_4$, the correct range-velocity estimate can be identified.

Although the acceleration of vehicles is utterly unpredictable in the traffic monitoring scenario, the acceleration of vehicles is limited. The maximum acceleration of high-performance cars is 5.4~m/s$^2$. For the linear motion model with constant acceleration 5.4~m/s$^2$, the predicted radar images at time $t_1$-$t_4$ are still significantly different between estimate~A and estimate~B, as shown in Fig.~6(c) and Fig.~6(d). The non-zero acceleration will induce changes in both $f_{\text{H}}$ and $f_\text{L}$. For estimate~A, the velocity changes slowly due to the limited acceleration. The range also changes slowly due to the low acceleration and low velocity (5~m/s) at time $t_0$. Therefore, the vehicle yields minor changes in both $f_{\text{H}}$ and $f_\text{L}$ at $t_1$-$t_4$ as shown in Fig.~6(c). For a vehicle with $R$ = 10~m, $v$ = 20~m/s, and $a$ = 5.4~m/s$^2$ at time $t_0$, enormous range changes due to high velocity induce major changes in $f_{\text{L}}$ at time $t_1$-$t_4$. Major changes in $f_{\text{L}}$ shift the peaks significantly, leading to a sparser peak pattern in Fig.~6(d) than the pattern in Fig.~6(c). Therefore, the velocity is crucial to filter out the incorrect estimates herein since the range change depends on the velocity. The limited acceleration in the traffic scenario is a minor factor in changing the peaks in the radar image.
\subsection{Discussions}
It has been mentioned that the signal processing complexity of applying 2D-DFT is computationally high. An OFDM sensing system with parameters in Table~II needs $2N\times{\mathcal{O}(N^2)}$ complex multiplications (480 DFTs in row and 480 IDFTs in column) to generate a range-velocity estimate. In contrast, the computational complexity of the proposed algorithm is relatively low because it avoids the DFT calculation both on all rows and all columns.

Since the phase shift of modulation signals along the frequency axis can extract the delay, and the phase shift of modulation signals along the time axis can extract the Doppler, the signaling overhead of the proposed diagonal structure is reduced by combining the phase shift along both the frequency axis and the time axis into a series of sensing signals along the diagonal of the resource block. Hence the range and velocity can be derived simultaneously. The sensing overhead of the conventional comb structure is $N^2/N^2_c$. The sensing overhead of the proposed diagonal structure is $N/N^2_c$, reducing the sensing overhead by a factor of $N$.

The only minor disadvantage of the proposed algorithm is that two unlabeled spectral peaks in the radar image cause ambiguity in the range-velocity estimates. The proposed multi-temporal data fusion method can be used to resolve ambiguity. Furthermore, steady objects do not create ambiguity by using the proposed algorithm. For example, Fig.~7 shows the radar image of a steady object with a range of 40~m and a velocity of 0~m/s. Another object with a range of 0~m and a velocity of 20~m/s produces the same radar image shown in Fig.~7. However, it directly contacts the antenna of the JCAS system due to the range $R$ being zero, which is kinematically infeasible. Therefore, the proposed algorithm can derive the range-velocity estimate of a steady object without ambiguity by using one diagonal block of sensing signals only.
\begin{figure}[t]
	\centerline{\includegraphics[width=0.9\linewidth, height=10cm, keepaspectratio]{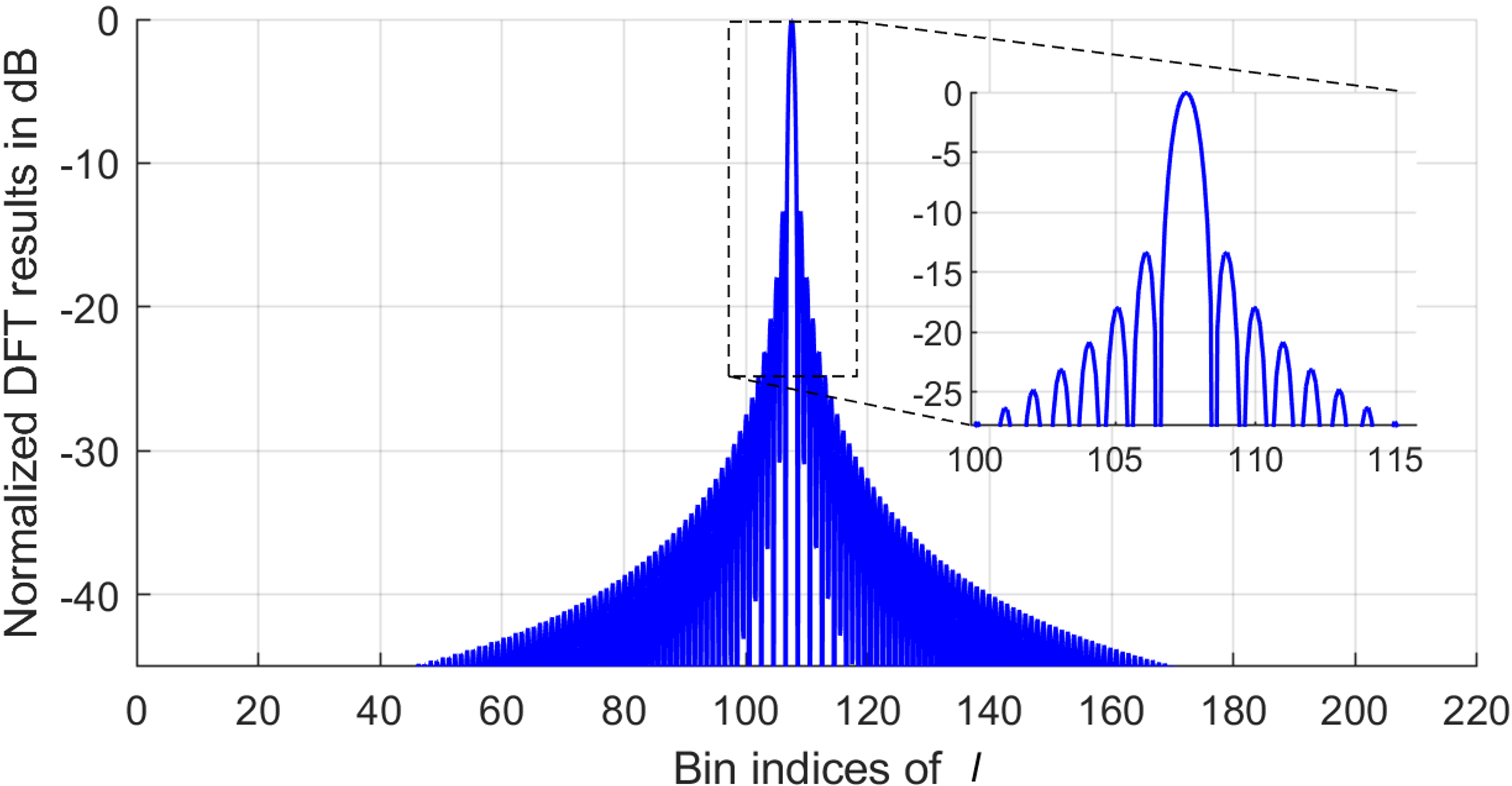}}
	\caption{The radar image of an object with range of 40~m and velocity of 0~m/s.}
	\label{fig_7}
\end{figure}

\section{Conclusions}
The typical processing algorithm of OFDM radar applies 2D-DFT for extracting the range and velocity information. This paper proposes a diagonal waveform structure and corresponding signal processing algorithm. With this approach, two advantages can be achieved. First, the signal processing algorithm is simple regarding computational complexity and memory requirement. Second, sensing overhead is significantly reduced by the proposed waveform structure. The minor disadvantage is that the proposed approach obtains the range and velocity in a coupled manner. Therefore, range-velocity ambiguity may occur. A multi-temporal data fusion method can be performed to resolve ambiguity. The simulations have proven the operability of the proposed waveform and signal processing algorithm.

\vspace{12pt}
\color{red}
\end{document}